\documentclass{aa}  

\usepackage{graphicx}
\usepackage{txfonts}
\usepackage{color,hyperref}
\definecolor{darkblue}{rgb}{0.0,0.0,0.3}
\hypersetup{colorlinks,breaklinks,
            linkcolor=blue,urlcolor=blue,
            anchorcolor=blue,citecolor=blue}

\begin{document}

\newcommand{\kms}{km s$^{-1}$}

\title{C$_2$H observations toward the Orion Bar}
\authorrunning{Z. Nagy et al.}

   \author{Z. Nagy\inst{1,2};
           V. Ossenkopf\inst{2};
           F. F. S. Van der Tak\inst{3,4};
           A. Faure\inst{5};
           Z. Makai\inst{2};
\and
           E. A. Bergin\inst{6}
          }

\institute{
Department of Physics and Astronomy, University of Toledo, 2801 West Bancroft Street, Toledo, OH 43606, USA \\
\email{zsofia.nagy.astro@gmail.com} 
\and
I. Physikalisches Institut, Universit\"at zu K\"oln, Z\"ulpicher Str. 77, 50937 K\"oln, Germany
\and
Kapteyn Astronomical Institute, University of Groningen, PO box 800, 9700 AV Groningen, The Netherlands 
\and
SRON Netherlands Institute for Space Research, Landleven 12, 9747 AD Groningen, The Netherlands
\and
Universit\'e Joseph Fourier/CNRS, Institut de Plan\'etologie et d'Astrophysique de Grenoble (IPAG) UMR 5274, 38041 Grenoble, France
\and
University of Michigan, Ann Arbor, MI 48197, USA
}

 
  \abstract
{
The ethynyl radical (C$_2$H) is one of the first radicals to be detected in the interstellar medium. Its higher rotational transitions have recently become available with the \textit{Herschel} Space Observatory.
}
{
We aim to constrain the physical parameters of the C$_2$H emitting gas toward the Orion Bar.
}
{We analyse the C$_2$H line intensities measured toward the Orion Bar CO$^+$ Peak and \textit{Herschel}/HIFI maps of C$_2$H, CH, and HCO$^+$, and a NANTEN map of [C{\sc{i}}]. We interpret the observed C$_2$H emission using the combination of \textit{Herschel}/HIFI and NANTEN data with radiative transfer and PDR models.}
{
Five rotational transitions of C$_2$H (from $N$=6$-$5 up to $N$=10$-$9) have been detected in the HIFI frequency range toward the CO$^+$ peak of the Orion Bar.
Based on the five detected C$_2$H transitions, a single component rotational diagram analysis gives a rotation temperature of $\sim$64 K and a beam-averaged C$_2$H column density of $4\times10^{13}$ cm$^{-2}$. 
The rotational diagram is also consistent with a two-component fit resulting in rotation temperatures of 43$\pm$0.2~K and 123$\pm$21~K, and beam-averaged column densities of $\sim8.3\times10^{13}$ cm$^{-2}$ and $\sim2.3\times10^{13}$ cm$^{-2}$ for the three lower-$N$ and for the three higher-$N$ transitions, respectively.
The measured five rotational transitions cannot be explained by any single parameter model.
According to a non-LTE model, most of the C$_2$H column density produces the lower$-N$ C$_2$H transitions and traces a warm ($T_{\rm{kin}}\sim100-150$ K) and dense ($n$(H$_2$)$\sim$10$^5$-10$^6$ cm$^{-3}$) gas. A small fraction of the C$_2$H column density is required to reproduce the intensity of the highest-$N$ transitions ($N$=9$-$8 and $N$=10$-$9) originating from a high density ($n$(H$_2$)$\sim$5$\times$10$^6$ cm$^{-3}$) hot ($T_{\rm{kin}}\sim400$ K) gas. The total beam-averaged C$_2$H column density in the model is 10$^{14}$ cm$^{-2}$.
A comparison of the spatial distribution of C$_2$H to those of CH, HCO$^+$, and [C{\sc{i}}] shows the best correlation with CH.}
{
Both the non-LTE radiative transfer model and a simple PDR model representing the Orion Bar with a plane-parallel slab of gas and dust suggest, that C$_2$H cannot be described by a single pressure component, unlike the reactive ion CH$^+$, which was previously analysed toward the Orion Bar CO$^+$ peak. 
The physical parameters traced by the higher rotational transitions ($N$=6-5,...,10-9) of C$_2$H may be consistent with the edges of dense clumps exposed to UV radiation near the ionization front of the Orion Bar.
}


\maketitle

\section{Introduction}

The ethynyl radical (C$_2$H) was one of the first radicals to be detected in the interstellar medium \citep{tucker1974}. It has been observed toward several types of regions, including diffuse clouds \citep{lucasliszt2000}, massive star-forming regions \citep{beuther2008}, photon-dominated regions (PDRs, e.g. \citealp{teyssier2004}) and dark clouds (e.g. \citealp{pratap1997}). 
C$_2$H is an important molecule in the carbon-chemistry, and is related to many C-bearing species. One example is the C$_3$H$^+$ ion, which has recently been detected toward the Horsehead PDR \citep{pety2012}. C$_2$H can also be an ingredient for the formation of carbon chain molecules, such as C$_5$, via a reaction with C$_3$. C$_2$H can also be converted to CO either by a neutral-neutral reaction with O, or by a reaction with carbon chain ions, such as HCO$^+$.  
Due to its role in the carbon chemistry, understanding the abundance of C$_2$H is important to understand the chemical network related to carbon.
The recent access to the higher rotational transitions provided by the \textit{Herschel} Space Observatory and the recently calculated inelastic collision rates \citep{spielfiedel2012} give an opportunity to probe the chemistry and excitation of C$_2$H.
Using the higher rotational transitions accessible by \textit{Herschel} and the available collision rates we probe the excitation of C$_2$H and the physical parameters that it traces toward the prototypical, high UV-illumination PDR, the Orion Bar. As photodissociation of larger carbon-chain molecules and PAHs is one of the possible formation routes of C$_2$H, PDRs provide a good source to study C$_2$H.

The Orion Bar is an ideal source to probe the excitation and chemistry of molecules in PDRs, due to its close distance of 414 pc \citep{menten2007} and its well-known structure and geometry. 
The Orion Bar is located between the Orion Molecular Cloud and an H{\sc{ii}} region illuminated by the Trapezium cluster. The FUV radiation field of the Trapezium cluster at the location of the Orion Bar is equivalent to $(1-4) \times 10^4 \chi_0$ in \citet{draine1978} units. 
Its orientation changes from face-on to nearly edge-on where the molecular emissions peak. The observations presented in this paper also correspond to the nearly edge-on orientation part of the Orion Bar. 
The geometrical enhancement of the column densities toward the nearly edge-on part of the Orion Bar was derived by multiple studies and is in the range between 4 and 20.
The tilt angle compared to a completely edge-on orientation was suggested to be 3$^\circ$ in the model of \citet{hogerheijde1995} and \citet{jansen1995}. A tilt angle of 3$^\circ$ is equivalent to an enhancement factor of 20 for the measured column densities. Based on O{\sc{i}} 1.317 $\mu$m emission \citet{walmsley2000} find a model that requires a geometrical enhancement factor of 5 to convert the observed column densities into face-on values. \citet{neufeld2006} found the geometrical enhancement factor to be 4 based on measured C$^+$ column densities. 
Using a clumpy 3D PDR model, \citet{andreelabsch2014} successfully reproduced the Orion Bar stratification using a clumpy edge-on cavity wall, and claim that a model of a convex filament fails to describe the structure of the Orion Bar.
The average kinetic temperature was estimated to be 85 K \citep{hogerheijde1995}. Closer to the ionization front, higher temperatures are also measured, for example OH transitions observed with \textit{Herschel}/PACS are consistent with 160-220 K gas \citep{goicoechea2011}, and CH$^+$ observations with temperatures around 500 K \citep{nagy2013}.
Part of the molecular line emission measured toward the Orion Bar corresponds to an 'interclump medium' with densities between a few 10$^4$ and 2$\times$10$^5$ cm$^{-3}$ \citep{simon1997}. Other molecular lines have been suggested to originate in clumps with densities in the range between 1.5$\times$10$^6$ and 6$\times$10$^6$ cm$^{-3}$ \citep{lisschilke2003}. 

Previous C$_2$H observations toward the Orion Bar cover the lower-$N$ rotational transitions. 
\citet{cuadrado2015} analyse the $N$=1-0,...,4-3 transitions. The $N$=4$-$3 transition was observed by \citet{hogerheijde1995}, \citet{jansen1995}, and \citet{vanderwiel2009}), and the $N$=1-0 transition by \citet{fuente1996}. 
In this paper we analyse emission from five higher rotational transitions to constrain the physical parameters of the C$_2$H emitting gas toward the Orion Bar.

\section{Observations and Data reduction}

The observations presented in this paper are part of the HEXOS\footnote{Herschel observations of EXtra-Ordinary Sources} guaranteed-time key program \citep{bergin2010} for the HIFI instrument \citep{degraauw2010} of the \textit{Herschel} Space Observatory \citep{pilbratt2010}.
The CO$^+$ peak ($\alpha_\mathrm{J2000}=\rm{05^h35^m20.6^s}$, $\delta_\mathrm{J2000}=-05^\circ 25'14''$) of the Orion Bar has been observed over the full HIFI range (480-1910 GHz) as a spectral scan. 
An overview of the full spectral line survey will be presented in a forthcoming paper (Nagy et al., 2015).
The data were reduced using HIPE \citep{ott2010} pipeline version 9.0 and 10.0 and are calibrated to $T^\star_A$ scale. 
The sideband deconvolution was done using the \textit{doDeconvolution} task in HIPE. 
The double sideband (DSB) scans were first deconvolved with the strongest lines ($T_{\rm{A}}^*$ $>$ 10 K) removed. This deconvolution results in a single sideband (SSB) spectrum with very little contribution from deconvolution ghosts, but without data at the frequencies of the strongest lines.
Therefore, we performed another deconvolution of the data including the strongest lines. The data around the frequencies of the strong lines was then used together with the first deconvolution result, providing a single sideband spectrum in the total observed frequency range. 
Finally, the strong lines were incorporated into the weak line SSB spectrum. 

The continuum at the observed frequencies is negligible, as it is similar to the measured rms noise level. 
The C$_2$H transitions detected in the HIFI survey are listed in Table \ref{gauss_param} including spectroscopic parameters from the Cologne Database for Molecular Spectroscopy (CDMS, \citealp{mueller2005}). The hyperfine structure is not resolved by HIFI, therefore the detected lines are blends of multiple hyperfine components. Table \ref{gauss_param} lists the spectroscopic parameters for the strongest hyperfine components.
The spectral resolutions at the frequencies of the observed transitions (from $N$=6-5 to $N$=10-9) are 0.29, 0.25, 0.21, 0.19, and 0.17 \kms.

To compare the line intensities of the transitions detected with different beam sizes in the HIFI line survey, we convert all the observed line intensities to a common $\sim$40.5$''$ resolution. We derive conversion factors between the original beam sizes and $\sim$40.5$''$ based on the integrated intensity map of the C$_2$H 4$-$3 transition from the James Clerk Maxwell Telescope (JCMT) Spectral Legacy Survey (\citealp{vanderwiel2009}). 
The derived correction factors between the beam sizes of the C$_2$H transitions and the 40.5$''$ beam are 0.94, 0.88, 0.84, and 0.79 from beam sizes of 34.7$''$, 30.4$''$, 27.0$''$, and 24.3$''$ , respectively.
Table \ref{gauss_param} includes the parameters that are measured with the original beam sizes. For the non-LTE models in Sect. \ref{nonlte} we use the values that have been corrected for the changing beam size.

In addition to the HIFI spectral scans, a $115'' \times 65''$ area centered on $\alpha_\mathrm{J2000}=\rm{05^h35^m20.81^s}$, $\delta_\mathrm{J2000}=-05^\circ 25'17.1''$ with a position angle of 145$^\circ$ was mapped in the $N$=6--5 transition of C$_2$H with HIFI, in on-the-fly (OTF) mapping mode with position-switch reference. The center of the maps is consistent with the CO$^+$ peak (it is 4$''$ off from the CO$^+$ peak) where the spectral scans are pointed. The CO$^+$ peak is close to the PDR surface and to the region where vibrationally excited H$_2$ peaks (e.g. \citet{walmsley2000} and Fig. 1. of \citet{nagy2013}).
The fully sampled map was pipelined with HIPE 11.1 and exported to CLASS for further analysis, including baseline subtraction using linear baselines. 
As a comparison to the C$_2$H emission, we use maps of the HCO$^+$ $J$=6-5 transition (535061.6 MHz) and of the CH $^2\Pi_{3/2}$ transitions around 536 GHz ($N$=1-0, $J$=3/2-1/2, $F$=$2^- - 1^+$ at 536.76115 MHz, $N$=1-0, $J$=3/2-1/2, $F$=$1^- - 1^+$ at 536782.0 MHz, $N$=1-0, $J$=3/2-1/2, $F$=$1^- - 0^+$ at 536.79568). These maps have the same parameters and were reduced with the same HIPE version and methods as the C$_2$H $N$=6-5 map. The spectral resolution of the observed HCO$^+$ and CH transitions is 0.28 \kms.

In addition to the HIFI maps, we use a [C{\sc{i}}] $^3$P$_1$--$^3$P$_0$ (492.1607 GHz) map observed with the NANTEN2-4m antenna, which is located at 4865 m altitude in Pampa la Bola in northern Chile. 
The observations were taken in October and November 2011 with a dual-channel 460/810 GHz SMART receiver. The receiver temperature was $\sim$250 K at the band center. 
The velocity resolution (or channel spacing) is 0.68 \kms at 460 GHz. 
The observed spectra were calibrated using the atmospheric model Atmospheric Transmision at Microwave \citep{pardo2001}. The half power beam width (HPBW) is $\sim$37$''$. 
The data were calibrated to main beam temperature using main beam and forward efficiencies of 0.50 and 0.86, respectively.

\section{The spatial distribution and velocity structure of C$_2$H}   

\begin{figure}[ht]
\centering
\includegraphics[height=10cm, angle=-90, trim=2.0cm 2.5cm 2cm 1cm,clip=true]{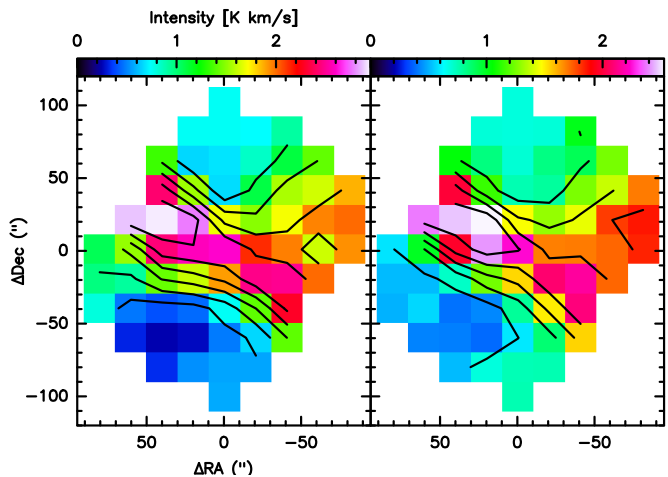}
\caption{Integrated intensities of the C$_2$H $N=6-5$ $J=13/2-11/2$ (left) and $J=11/2-9/2$ (right) doublet. The contour levels go from 0.2 to 2.8 K \kms spaced by 0.4 K \kms.}
\label{orionbar_c2h_int}
\end{figure}

Figure \ref{orionbar_c2h_int} shows the integrated intensity of the C$_2$H $N=6-5$ $J=13/2-11/2$ and $J=11/2-9/2$ doublet. C$_2$H emission has been detected toward the Bar and also perpendicular to the Bar, corresponding to the Orion Ridge. The C$_2$H $N$=6--5 emission is not as extended as other species including CH and [C{\sc{i}}] (Fig. \ref{orionbar_int}). The C$_2$H integrated intensity does not peak at the CO$^+$ peak (the center of the maps), but toward the north-eastern part of the Bar covered by the C$_2$H map, around offsets (45$''$, 25$''$). 
The difference in the position where the C$_2$H and CO$^+$ line intensities peak is likely to be a result of different physical parameters toward the two positions. As mentioned in Appendix \ref{form_dest}, C$_2$H formation for the Orion Bar is expected to be dominated by a reaction of C$_2$ with H$_2$. CO$^+$ is produced in a reaction between C$^+$ and OH, where OH is produced from O and H$_2$.
Therefore, the CO$^+$ and C$_2$H peaks likely have different H$_2$ volume densities. The region where C$_2$H peaks overlaps with one of the OH$^+$ intensity peaks (at a velocity of $\sim$12 \kms, \citealp{vandertak2013}), which lies about 10$''$ south-east of the CO 10-9 intensity peak at a velocity of $\sim$12 \kms, which is observed toward the H$^{13}$CN clumps 2 and 3 identified by \citet{lisschilke2003}.

\begin{table*}[!ht]
\begin{minipage}[!h]{\linewidth}\centering
\caption{Spectroscopic and Gaussian fit parameters of the C$_2$H lines observed toward the CO$^+$ peak of the Orion Bar.}
\label{gauss_param}
\renewcommand{\footnoterule}{}
\renewcommand{\thefootnote}{\alph{footnote}}
\begin{tabular}{llrrrrlrrr}
\hline
Transition& $A_{\rm{ij}}$& $E_u$& $\nu$ & $\int T_{\rm MB}{\mathrm d}V$ & $V_{\rm LSR}$ & $\Delta V$    & $T_{\rm peak}$& rms& $\Theta_{\rm{B}}$\\
& (s$^{-1}$)& (K)& (MHz)& (K km s$^{-1}$)& (km s$^{-1}$)& (km s$^{-1}$)& (K)& (K)& ($''$)\\
\hline

$N=6-5$, $J=13/2-11/2$& $4.54\times10^{-4}$& 88.0& 
523971.6& 3.64$\pm$0.03& 10.59$\pm$0.08& 2.65$\pm$0.03& 1.29& 0.02& 40.5\\
                          
$N=6-5$, $J=11/2-9/2$& $4.46\times10^{-4}$& 88.0& 
524033.9& 2.95$\pm$0.03& 10.58$\pm$0.01& 2.61$\pm$0.03& 1.06& 0.02& 40.5\\

$N=7-6$, $J=15/2-13/2$& $7.30\times10^{-4}$& 117.4& 
611267.2& 2.38$\pm$0.16& 10.77$\pm$0.07& 2.44$\pm$0.19& 0.91& 0.13& 34.7\\

$N=7-6$, $J=13/2-11/2$& $7.21\times10^{-4}$& 117.4& 
611329.7& 2.06$\pm$0.14& 10.70$\pm$0.03& 2.50$\pm$0.20& 0.78& 0.11& 34.7\\

$N=8-7$, $J=19/2-17/2$& $1.10\times10^{-3}$& 150.9&
698544.8& 1.32$\pm$0.09& 10.67$\pm$0.07& 2.12$\pm$0.17& 0.59& 0.09& 30.4\\

$N=8-7$, $J=17/2-15/2$& $1.09\times10^{-3}$& 150.9&
698607.5& 1.23$\pm$0.10& 10.67$\pm$0.08& 2.23$\pm$0.21& 0.52& 0.10& 30.4\\

$N=9-8$, $J=19/2-17/2$& $1.58\times10^{-3}$& 188.6& 
785802.1& 1.21$\pm$0.09& 10.35$\pm$0.09& 2.51$\pm$0.23& 0.45& 0.08& 27.0\\

$N=9-8$, $J=17/2-15/2$& $1.57\times10^{-3}$& 188.6&
785865.0& 0.96$\pm$0.10& 10.56$\pm$0.12& 2.44$\pm$0.28& 0.37& 0.09& 27.0\\

$N=10-9$, $J=21/2-19/2$& $2.18\times10^{-3}$& 230.5& 
873036.4& 0.82$\pm$0.10& 11.12$\pm$0.15& 2.23$\pm$0.30& 0.35& 0.11& 24.3\\

$N=10-9$, $J=19/2-17/2$& $2.16\times10^{-3}$& 230.5& 
873099.5& 1.25$\pm$0.11& 10.11$\pm$0.18& 3.00$\pm$0.50& 0.30& 0.09& 24.3\\

\hline
\end{tabular}
\end{minipage}
\end{table*}

\begin{figure}[ht]
\centering
\includegraphics[width=7.5cm, angle=0, trim=0cm 0cm 0cm 0cm,clip=true]{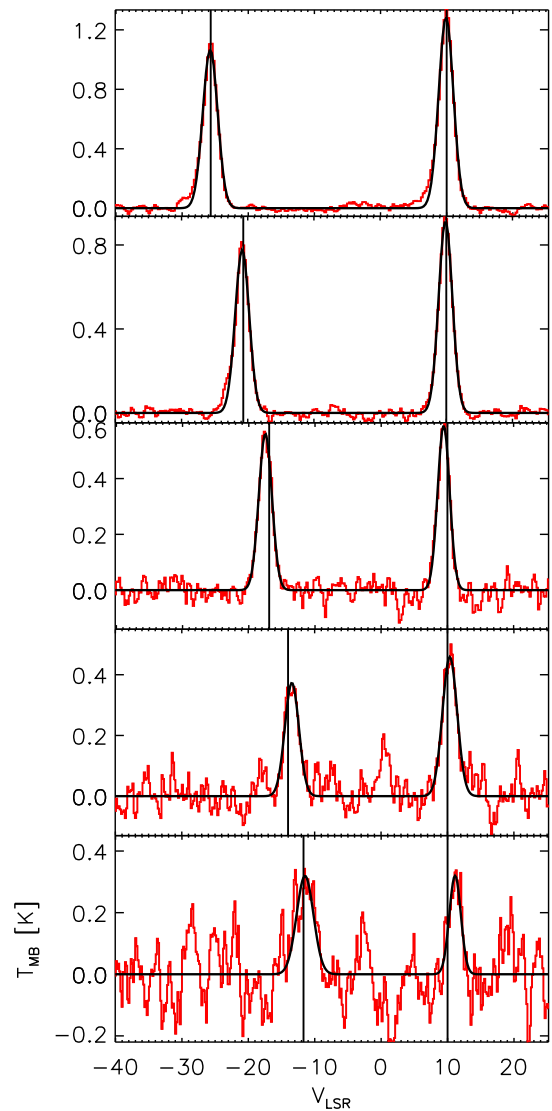}
\caption{Line profiles of the C$_2$H doublets observed toward the CO$^+$ peak of the Orion Bar, from $N$=6-5 (top) to $N$=10-9 (bottom). The Gaussian fits shown in Table \ref{gauss_param} are overlaid on the profiles. The black lines show the expected LSR velocity of 10 \kms for both components.}
\label{line_profiles_c2h}
\end{figure}

\begin{figure*}[ht]
\centering
\includegraphics[width=13cm, angle=0, trim=0cm 0cm 0cm 0cm,clip=true]{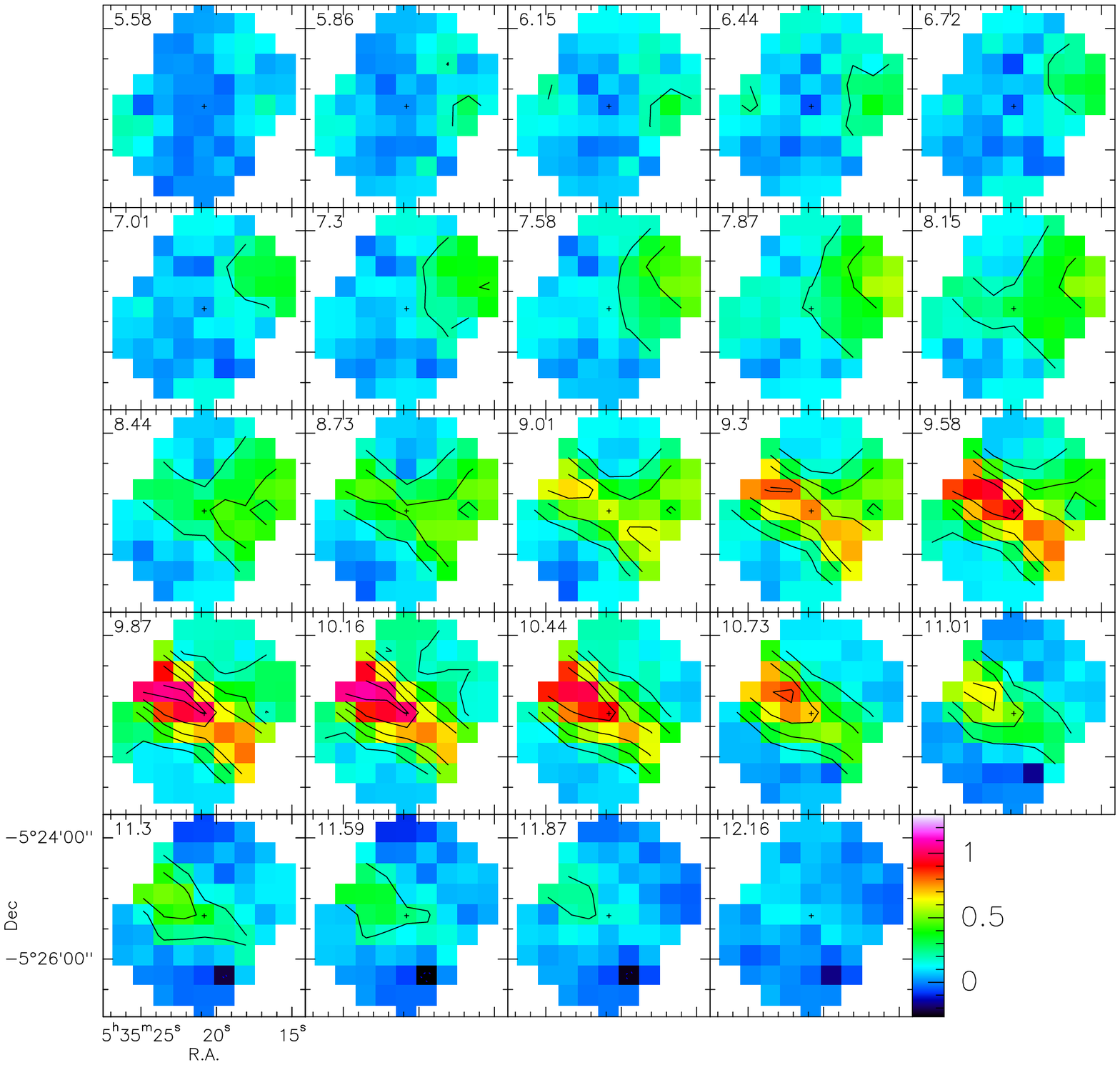}
\caption{Velocity channel maps of the C$_2$H $N=6-5$ $J=13/2-11/2$ transition toward the Orion Bar.}
\label{chanmap_c2h}

\centering
\includegraphics[width=8.0cm, angle=-90, trim=8.0cm -1.0cm 2.0cm 0.0cm,clip=true]{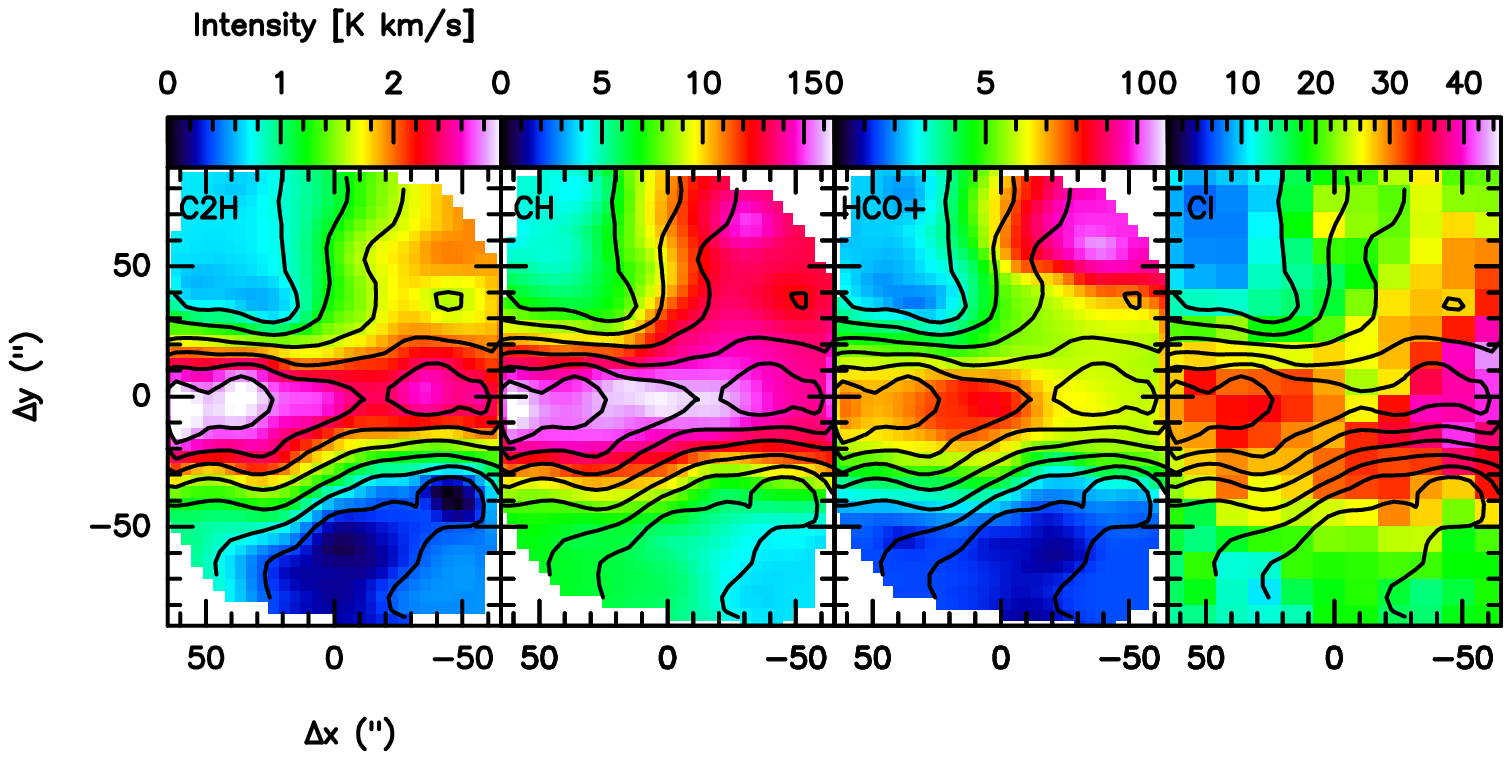}
\caption{Integrated intensity maps of C$_2$H $N$=6-5, CH $^2\Pi_{3/2}$, HCO$^+$ $J$=6-5, and [C{\sc{i}}] $^3$P$_1$--$^3$P$_0$ toward the Orion Bar. The C$_2$H integrated intensities are over-plotted on each map (black contours). The maps have been rotated by -35$^\circ$.}
\label{orionbar_int}
\end{figure*}

Apart from the $N$=6-5 doublet, four more doublets (from $N$=7-6 to $N$=10-9) have been detected toward the CO$^+$ peak of the Orion Bar, which the $N$=6-5 map is centered on.
Figure \ref{line_profiles_c2h} shows the line profiles of the C$_2$H doublets observed there. The observed two lines, which are blends of hyperfine transitions, are shifted with $\sim$62 MHz in every case, which is equivalent to different velocity intervals due to the change in the frequency of the transitions. Table \ref{gauss_param} lists the line intensities of these C$_2$H transitions as obtained by Gaussian fitting. 
The observed doublet intensity ratios are 1.23, 1.16, 1.07, 1.26, and 0.66 for $N$=6-5,...,10-9, respectively.
In LTE, the doublet ratio is given by the ratio of the A coefficients so that we expect that the blue-shifted component to be about 2\% brighter than the red one. The opposite ratio in $N$=10-9 is probably due to the lower S/N leading to an overestimate of the width and integrated intensity in the red-shifted component.
The peak velocities are in the range between 10.6-11.1 \kms. 
The variation in the peak velocities falls slightly above the uncertainty from the frequency accuracy of the HIFI products and the fit accuracy so that it may indicate a small velocity variation within the C$_2$H emitting gas, but it is very small compared to the line width, so that we can ignore it here, still being sure to analyse the same gas in all transitions.
As shown in Fig. \ref{chanmap_c2h} on the velocity channel maps of C$_2$H, the Orion Ridge is most prominent at velocities lower than 9 \kms. Most C$_2$H emission toward the Orion Bar is seen at velocities in the range between 9 and 11 \kms. 
The velocity structure of C$_2$H is similar to that seen in OH$^+$ \citep{vandertak2013}: the blue-shifted part of the emission peaks toward the Orion Ridge, emission around the expected source velocity (10 \kms) has contributions both from the Orion Bar and the Orion Ridge, while emission corresponding to the red-shifted velocities only comes from the Bar.
However, the peak velocities at which the Orion Bar and Ridge are detected are slightly shifted for C$_2$H compared to those for OH$^+$. OH$^+$ at velocities of 9--10 \kms shows emission mainly toward the Orion Ridge. In the same velocity interval C$_2$H emission peaks toward the Orion Bar, and shows little emission toward the Orion Ridge. 
This could be explained by low-density gas in the Orion Ridge at $v < 9$~\kms{} providing a lower
C$_2$H excitation compared to OH$^+$ for that component.

\begin{figure}[ht]
\centering
\includegraphics[width=8.0cm, angle=0, trim=0cm 0.5cm 0cm 21cm,clip=true]{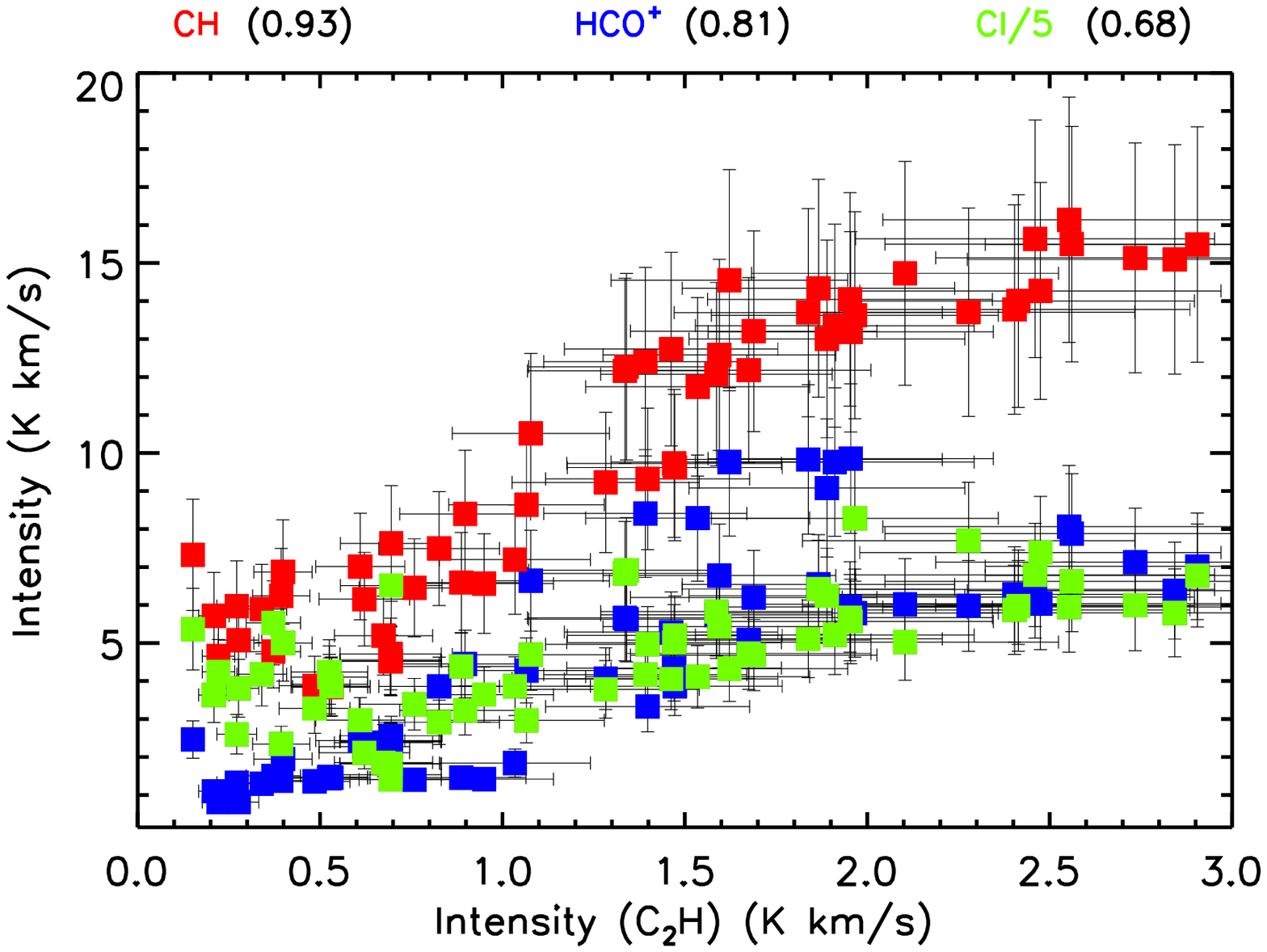}
\caption{Integrated line intensity correlations between the C$_2$H $N=6-5$ $J=13/2-11/2$ line and HCO$^+$, CH, and [C{\sc{i}}]. The corresponding correlation coefficients are shown in parentheses.}
\label{c2h_corr}

\includegraphics[width=8.0cm, angle=0, trim=0cm 0.0cm 0cm 0cm,clip=true]{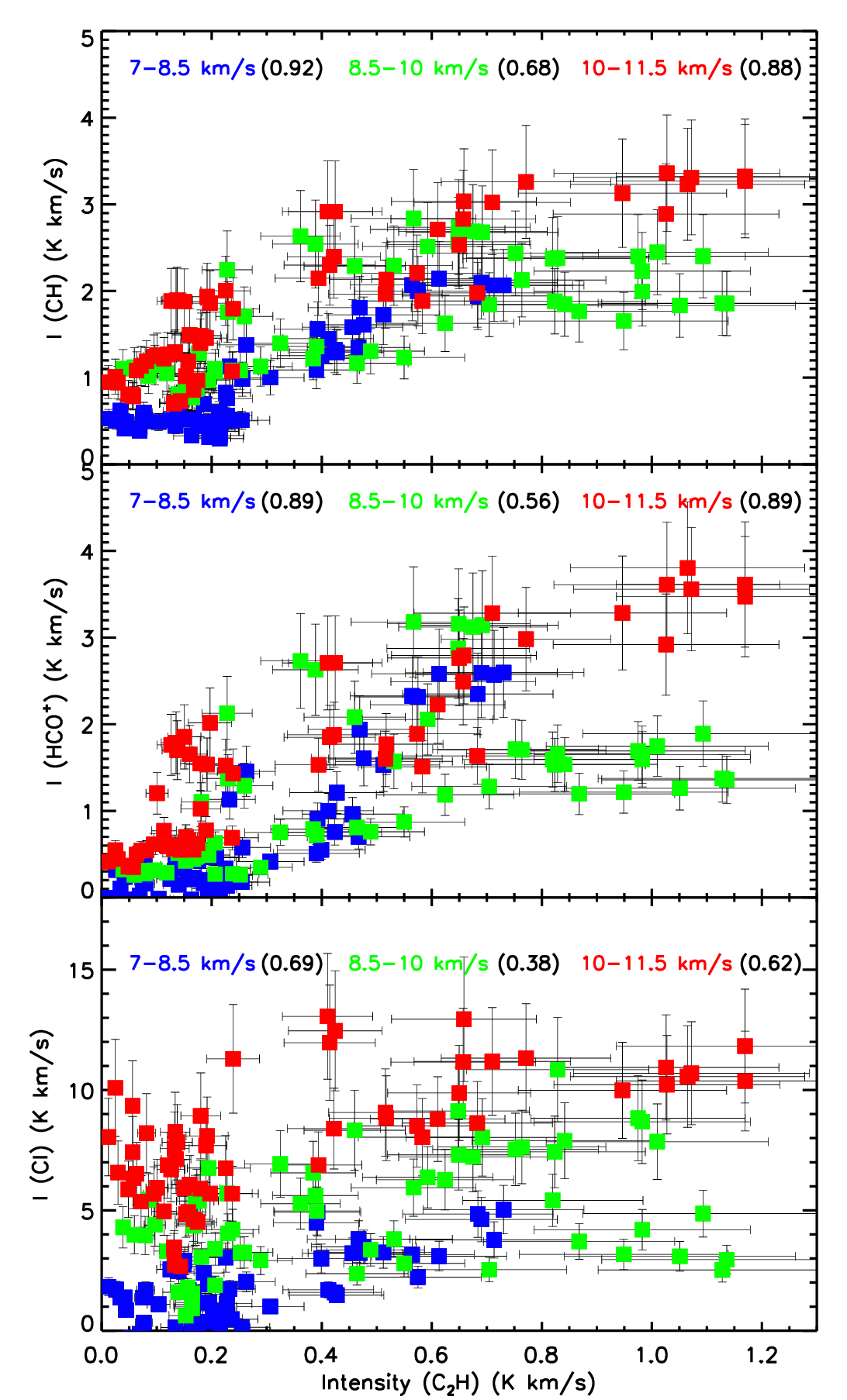}
\caption{Line intensity spatial correlations between the C$_2$H $N=6-5$ $J=13/2-11/2$ line and HCO$^+$, CH, and [C{\sc{i}}]. The corresponding correlation coefficients are shown in parentheses.}
\label{c2h_corr_intervals}
\end{figure}

Figure \ref{orionbar_int} shows the spatial distribution of C$_2$H $N$=6-5 and lines of other species that are expected to trace regions relatively close to the PDR surface (equivalent to low visual extinctions): CH $^2\Pi_{3/2}$ $N$=1-0 $J$=3/2-1/2, HCO$^+$ $J$=6-5, and [C{\sc{i}}] $^3$P$_1$-$^3$P$_0$. 
Out of the four line emissions shown in Fig. \ref{orionbar_int}, the lines of CH and [C{\sc{i}}] show the largest spatial extent over the area analysed in this paper. All these lines peak at a similar position toward the Orion Bar. 
HCO$^+$ $J$=6-5 emission toward the Orion Bar is less extended compared to the lines of the other molecules, especially C{\sc{i}} and CH.
HCO$^+$ $J$=6-5 is also the only line in Fig. \ref{orionbar_int} that peaks toward the Orion Ridge and not toward the Bar.
The comparison of the spatial distribution of C$_2$H to those of HCO$^+$, CH, and [C{\sc{i}}] based on the line emissions shown in Fig. \ref{orionbar_int} is straightforward, due to the similar beam size of the observed transitions. In the following paragraphs while comparing the spatial distribution of C$_2$H to those of CH, HCO$^+$, and [C{\sc{i}}], we refer to the transitions shown in Fig. \ref{orionbar_int}. The spatial distribution of the species may vary depending on the transitions used.
The comparison of the spatial distributions can be converted to correlation plots (Fig. \ref{c2h_corr}). C$_2$H shows the strongest correlation with CH, compared to HCO$^+$ and [C{\sc{i}}]. 
The corresponding correlation coefficients for CH, HCO$^+$, and [C{\sc{i}}] are 0.93, 0.81, and 0.68, respectively.

As seen on the velocity channel maps, the velocity range covered by the observed lines corresponds to at least two regions: the Orion Bar and the Orion Ridge. 
They represent two different PDRs with different conditions so that we expect a different behaviour.
Therefore, the spatial correlation may be different for the spatially different regions. Fig. \ref{c2h_corr_intervals} shows spatial correlations between C$_2$H, CH, HCO$^+$, and [C{\sc{i}}] calculated in three velocity intervals: 7--8.5 \kms, 8.5--10 \kms, and 10--11.5 \kms. The blue-shifted velocity interval is dominated by emission from the Orion Ridge. The central velocity interval has contributions both from the Orion Bar and the Orion Ridge. The red-shifted velocity interval is dominated by emission from the Orion Bar. 
The best correlation is found for the blue-shifted (correlation coefficients of 0.69--0.92), and the worst for the intermediate velocity intervals (correlation coefficients of 0.38--0.68). The decrease of the correlation coefficient for the central velocity interval may be related to optical depth effects.
Based on these results, CH and HCO$^+$ may be good tracers of C$_2$H for the Orion Ridge. 
Taking into account the red-shifted velocity interval only, these species may probe a similar gas to C$_2$H.
Among the molecules studied here, the gas traced by [C{\sc{i}}] is the least related to the gas traced by C$_2$H both for the Bar and the Ridge.

\section{C$_2$H column densities - single excitation temperature}
\label{coldens_lte}

One method to estimate the C$_2$H column density is interpreting the intensities of the observed five doublets by a rotational diagram. The observed integrated intensities can be directly converted to upper state column densities by assuming that the transitions are optically thin, the level populations can be characterized by a single excitation temperature, and that the line emission has a uniform beam filling. 
In the rotational diagram method the measured integrated main-beam temperatures of lines ($\int T_{\mathrm{MB}} \mathrm{d}V$ K km s$^{-1}$) are converted to the column densities of the molecules in the upper level ($N_\mathrm{u}$) using:
\begin{equation}
	\label{rotdiagram}
		\frac{N_\mathrm{u}}{g_\mathrm{u}}=\frac{N_{\mathrm{tot}}}{Q(T_{\mathrm{rot}})} \exp\left({-\frac{E_\mathrm{u}}{T_{\mathrm{rot}}}}\right)=\frac{1.67 \cdot 10^{14}}{\nu \mu^2 S} \int T_{\mathrm{MB}} \mathrm{d}V,
\end{equation}
with $g_\mathrm{u}$ the statistical weight of level u, $N_{\mathrm{tot}}$ the total column density in cm$^{-2}$, $Q(T_{\mathrm{rot}})$ the partition function for $T_{\mathrm{rot}}$, $E_\mathrm{u}$ the upper level energy in K, $\nu$ the frequency in GHz, $\mu$ the permanent dipole moment in Debye and $S$ the line strength value. A linear fit to $\ln(N_\mathrm{u}/g_\mathrm{u})$ - $E_\mathrm{u}$ gives $T_\mathrm{rot}$ as the inverse of the slope, and $N_\mathrm{tot}$, the column density as the intercept. The rotational temperature would be expected to be equal to the kinetic temperature if all levels were thermalized.

\begin{figure}[ht]
\centering
\includegraphics[width=6.5cm, angle=-90, trim=0cm 0cm 0cm 0cm,clip=true]{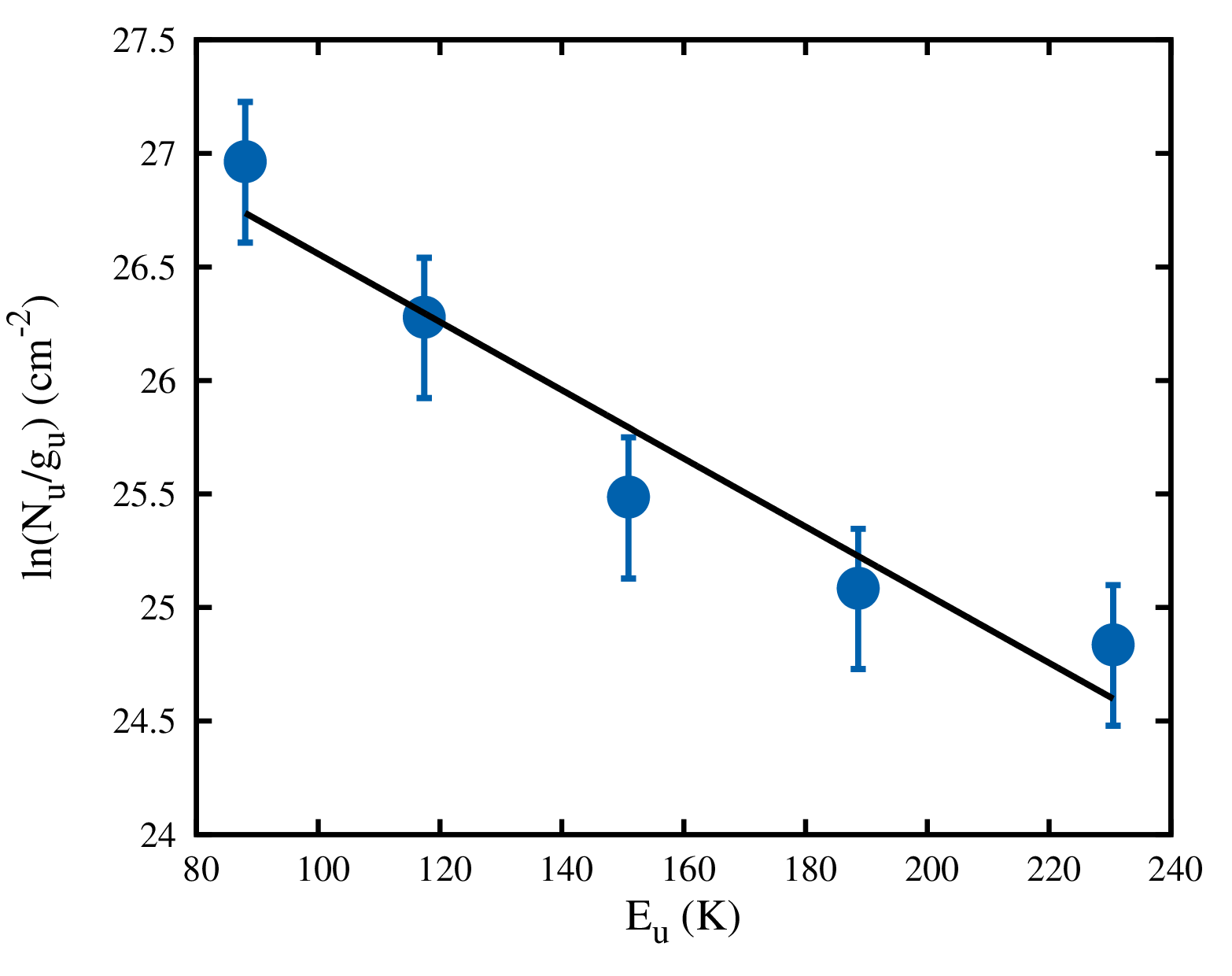}
\caption{The rotational diagram of C$_2$H, corresponding to a rotational temperature of 64$\pm$11 K and a column density of $(4 \pm 0.02) \times 10^{13}$ cm$^{-2}$.}
\label{c2h_rotdiagram}
\end{figure}

Fig. \ref{c2h_rotdiagram} shows the resulting rotation diagram when adding the hyperfine components of each rotational transition, which corresponds to a rotation temperature of $\sim$64$\pm$11 K and a C$_2$H column density of $(4 \pm 0.7) \times 10^{13}$ cm$^{-2}$. The ratios of the intensities of the hyperfine components are consistent with the optically thin assumption.

The curved rotational diagram in Fig. \ref{c2h_rotdiagram} may be an evidence for multiple density and temperature components to be present in the region covered by the HIFI beam. 
Therefore, we fitted the $N$=6$-$5,...,8$-$7 and the $N$=8$-$7,...,10$-$9 transitions separately as well. The rotation diagram of the lower-$N$ transitions gives a rotation temperature of 43$\pm$0.2~K and a column density of $(8\pm0.03)\times10^{13}$ cm$^{-2}$. 
The rotation diagram fit for the higher-$N$ transitions gives a rotation temperature of $\sim$123$\pm$21~K and a column density of $(2\pm0.4)\times10^{13}$ cm$^{-2}$.
In case of the two-component rotational diagram, the total beam-averaged C$_2$H column density is 10$^{14}$ cm$^{-2}$. This is similar to the maximum value that \citet{jansen1995} found along their observed cut toward the Orion Bar, and factors of 2-5 below the values that \citet{vanderwiel2009} derived, both corresponding to beam sizes of $\sim$15$''$.

\section{C$_2$H column densities - non-LTE calculation}
\label{nonlte}

As inelastic collision rates for C$_2$H recently became available, a non-LTE analysis is also possible. Non-LTE excitation is expected to be important as the transitions studied here have critical densities above the average $\sim$10$^5$ cm$^{-3}$ Orion Bar H$_2$ volume density \citep{hogerheijde1995}. For example, assuming a gas temperature of 100~K, for C$_2$H-H$_2$ collisions the $N$=6-5, $J$=13/2-11/2 transition have a critical density of 6.8$\times$10$^7$ cm$^{-3}$, and the $N$=8-7, $J$=17/2-15/2 a critical density of 1.6$\times$10$^8$ cm$^{-3}$.
While the LTE approach provides information on the C$_2$H column density and excitation temperature, the non-LTE analysis also gives an estimate on the kinetic temperature and H$_2$ volume density of the C$_2$H emitting gas. 
This is, of course, only possible if the excitation deviates from LTE, but this is expected at least for the high-$N$ transitions with very large critical densities.
C$_2$H-He collision rates were calculated by \citet{spielfiedel2012}, which we scaled to represent C$_2$H-H$_2$ collisions following \citet{schoier2005}, by a factor of 1.36, based on the ratio of the reduced mass of the C$_2$H-He and C$_2$H-H$_2$ systems. These collision rates are available for energies up to 104.9 cm$^{-1}$, covering three rotational transitions analysed in this paper ($N$=6$-$5, 7$-$6, and 8$-$7). The rates are calculated for a temperature range between 5 and 100 K. We extrapolated these collision rates for temperatures up to 400 K and energies up to $\sim$445 cm$^{-1}$ (see Appendix \ref{extrapol}). Collisional rates for these higher levels and temperatures are clearly required and should be computed properly, in order to avoid extrapolations.
We also use collision rates with electrons, calculated in the Born approximation with Infinite Order Sudden (IOS) recoupling, calculated for temperatures in the range between 10 and 1000 K (see Appendix \ref{elec}).
We ran several RADEX \citep{vandertak2007} models with physical parameters that are expected for the Orion Bar, i.e. kinetic temperatures between 50 and 400 K and H$_2$ volume densities between 5$\times$10$^4$ cm$^{-3}$ and $10^7$ cm$^{-3}$.  
We also use a background radiation field based on \citet{arab2012}, which is a modified blackbody distribution with a dust temperature of $T_{\rm{d}}$=50 K and a dust emissivity index of $\beta$=1.6.
The used electron density has very little effect on the line intensities, which is a consequence of the small C$_2$H dipole moment (Appendix \ref{elec}).

\begin{figure}[ht]
\centering
\includegraphics[width=9cm, angle=0, trim=0cm -0.5cm 0cm 0cm,clip=true]{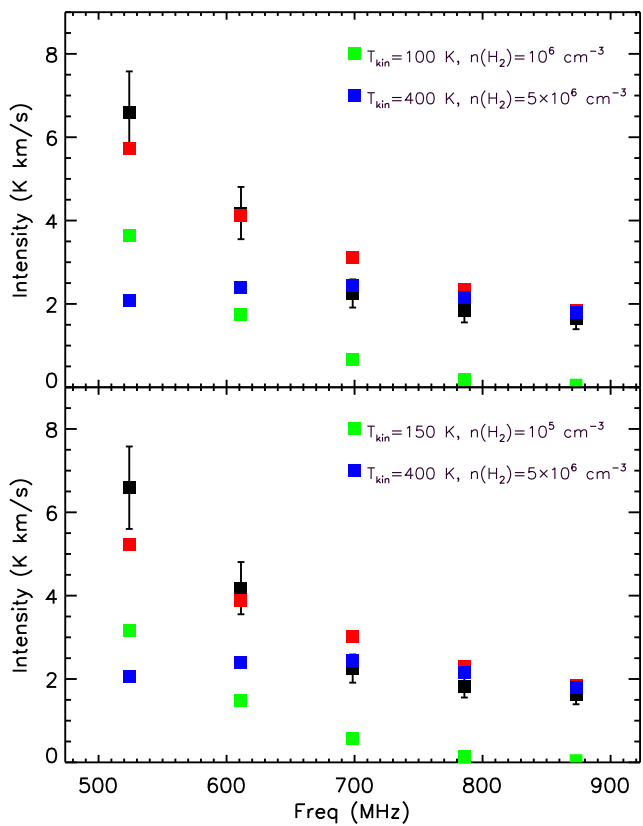}
\caption{Non-LTE C$_2$H line intensity predictions with RADEX vs the observed line intensities (black symbols). 
Top panel: The models corresponds to parameters of $n$(H$_2$)=10$^6$ cm$^{-3}$, $T_{\rm{kin}}$=100~K, $N$(C$_2$H)=$8\times10^{13}$ cm$^{-2}$ (green symbols), and to $n$(H$_2$)=5$\times$10$^6$ cm$^{-3}$, $T_{\rm{kin}}$=400~K, $N$(C$_2$H)=$2\times10^{13}$ cm$^{-2}$ (blue symbols). The sum of the two components is given by the red symbols.
Bottom panel: The models corresponds to parameters of $n$(H$_2$)=10$^5$ cm$^{-3}$, $T_{\rm{kin}}$=150~K, $N$(C$_2$H)=$8\times10^{13}$ cm$^{-2}$ (green symbols), and to $n$(H$_2$)=5$\times$10$^6$ cm$^{-3}$, $T_{\rm{kin}}$=400~K, $N$(C$_2$H)=$2\times10^{13}$ cm$^{-2}$ (blue symbols). The sum of the two components is given by the red symbols.
}
\label{c2h_radex}
\end{figure}

The observed line intensities are not consistent with C$_2$H originating in a single gas component toward the Orion Bar CO$^+$ peak. A two-component model is required to fit the observed C$_2$H intensities.
In this model the C$_2$H column density is dominated by a gas component with a temperature of 100-150~K, an H$_2$ volume density of 5$\times$10$^5$-$10^6$ cm$^{-3}$, and a C$_2$H column density of $8\times10^{13}$ cm$^{-2}$. To reproduce the integrated intensity of the highest-$N$ ($N$=9-8 and 10-9) doublets, a higher density ($5\times10^6$ cm$^{-3}$) hot ($T\sim400$ K) gas component is required. This second component has a column density of $2\times10^{13}$ cm$^{-2}$. 
The sum of the line intensities predicted by this two-component model are over-plotted on the observed line intensities in Fig. \ref{c2h_radex}.
This model is consistent with the $N$=6-5, 7-6, and 10-9 doublets within the observational errors, and over-predict the 8-7 and 9-8 doublets by 40\% and 30\%, respectively.
This model is also consistent with the assumption that the analysed C$_2$H transitions are optically thin: the opacity of every transition is less than 0.03 in each model.
The doublet ratios are well reproduced (13\% or better) by both two-component models for the $N$=6-5,...,9-8 transitions. The doublet ratio for the $N$=10-9 transition is reproduced within a factor of two.
The doublet ratios estimated with RADEX for the two-component model with n(H$_2$)=10$^6$ cm$^{-3}$, $T_{\rm{kin}}$=100 K and n(H$_2$)=$5\times10^6$ cm$^{-3}$ and $T$=400 K are 1.32, 1.27, 1.23, 1.22, and 1.22 for $N$=6-5,...,10-9, respectively. The doublet ratios for the two component model with n(H$_2$)=5$\times$10$^5$ cm$^{-3}$, $T_{\rm{kin}}$=150 K and n(H$_2$)=$5\times10^6$ cm$^{-3}$ and $T$=400 K are 1.28, 1.23, 1.20, 1.21, and 1.22.

The evidence of at least two different gas components observed toward the CO$^+$ peak is consistent with the clumpy picture of PDRs. Different density and temperature components are expected to be covered by the HIFI beam.
The existence of such high-density gas toward the Orion Bar CO$^+$ peak is consistent with what was found for OH \citet{goicoechea2011}.

The non-LTE models suggest a total C$_2$H column density of $\sim$10$^{14}$ cm$^{-2}$, dominated by the lowest-$N$ transitions. The models also suggest the evidence of a very high density gas component, that contributes only about 20\% of the C$_2$H column density, but is required to fit the highest-$N$ transitions.
The hot and dense gas that C$_2$H traces toward the Orion Bar based on the transitions analysed in this paper may be consistent with the UV-illuminated edges of dense clumps close to the ionization front of the PDR.

\section{Comparison to PDR models}

In earlier works based on the Orion Bar HIFI line survey, we presented an isobaric PDR model, which explained the observed CH$^+$ \citep{nagy2013} and OH$^+$ \citep{vandertak2013} emission toward the CO$^+$ peak of the Orion Bar. This model was computed using the 1.4.4 version of the Meudon PDR code \citep{lepetit2006} and corresponds to a pressure of $P=10^8$ cm$^{-3}$ K and represents a plane-parallel slab of gas and dust illuminated from two sides by a radiation field of $\chi=10^4$ on the front, and by $\chi$/1000 on the back side. We have implemented the collision rates used in Sect. \ref{nonlte} in order to compute integrated intensities for the observed transitions. 

\begin{figure}[ht]
\centering
\includegraphics[width=9.0cm, angle=0, trim=0cm 0cm 0cm 0cm,clip=true]{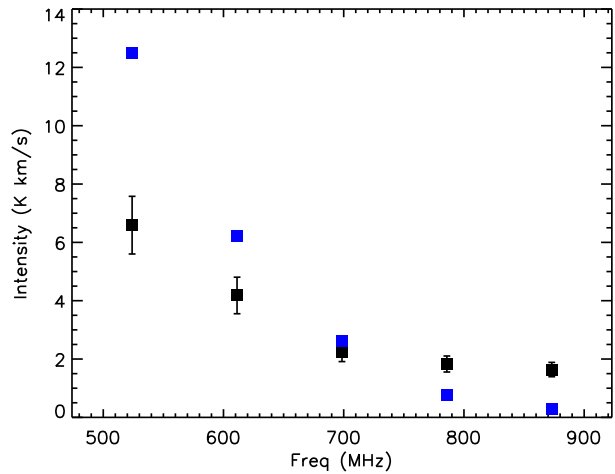}
\caption{C$_2$H line intensity predictions from a a PDR model with a pressure of $P=10^8$ cm$^{-3}$ K and a radiation field of $\chi=10^4$ (blue symbols) overplotted on the observations (black symbols).}
\label{c2h_int_meudoncode}
\end{figure}

Figure \ref{c2h_int_meudoncode} shows the calculated C$_2$H integrated intensities corresponding to the model, over-plotted on the observed integrated intensities. The model over-predicts the intensity of the $N$=6$-$5, $N$=7$-$6, and $N$=9$-$8 doublets by about a factors of two, 1.5, and 20\%. The model under-predicts the intensities of the $N$=10$-$9 and $N$=11$-$10 doublets by factors of 2.4 and 5.5. Even though the model predictions are close to most of the observed line intensities, it is clear that assuming a single pressure component is simplification, as the structure of PDRs is clumpy. The offset between the observed and modelled values may be explained by different gas components or clumps covered by the HIFI beam. 
The model predicts a face-on column density of 1.7$\times$10$^{14}$ cm$^{-2}$. Assuming a geometrical enhancement factor of four (e.g. \citealp{neufeld2006}), this is equivalent to an edge-on column density of 6.8$\times$10$^{14}$ cm$^{-2}$. This value is about a factor of 7 above the values predicted by the two-component rotational diagram and RADEX models. Despite the higher column density compared to the rotational diagram and RADEX models this model under-predicts the intensity of the highest-$N$ transitions. This confirms the result that the higher-$N$ transitions contribute only a minor part of the total C$_2$H column density.
The difference between the observed and modelled C$_2$H intensities suggests that most C$_2$H originates in a different gas component than CH$^+$, which was used to constrain the PDR model.

\begin{figure}[ht]
\centering
\includegraphics[width=4.7cm, angle=-90, trim=0cm 0.4cm 0cm 0cm,clip=true]{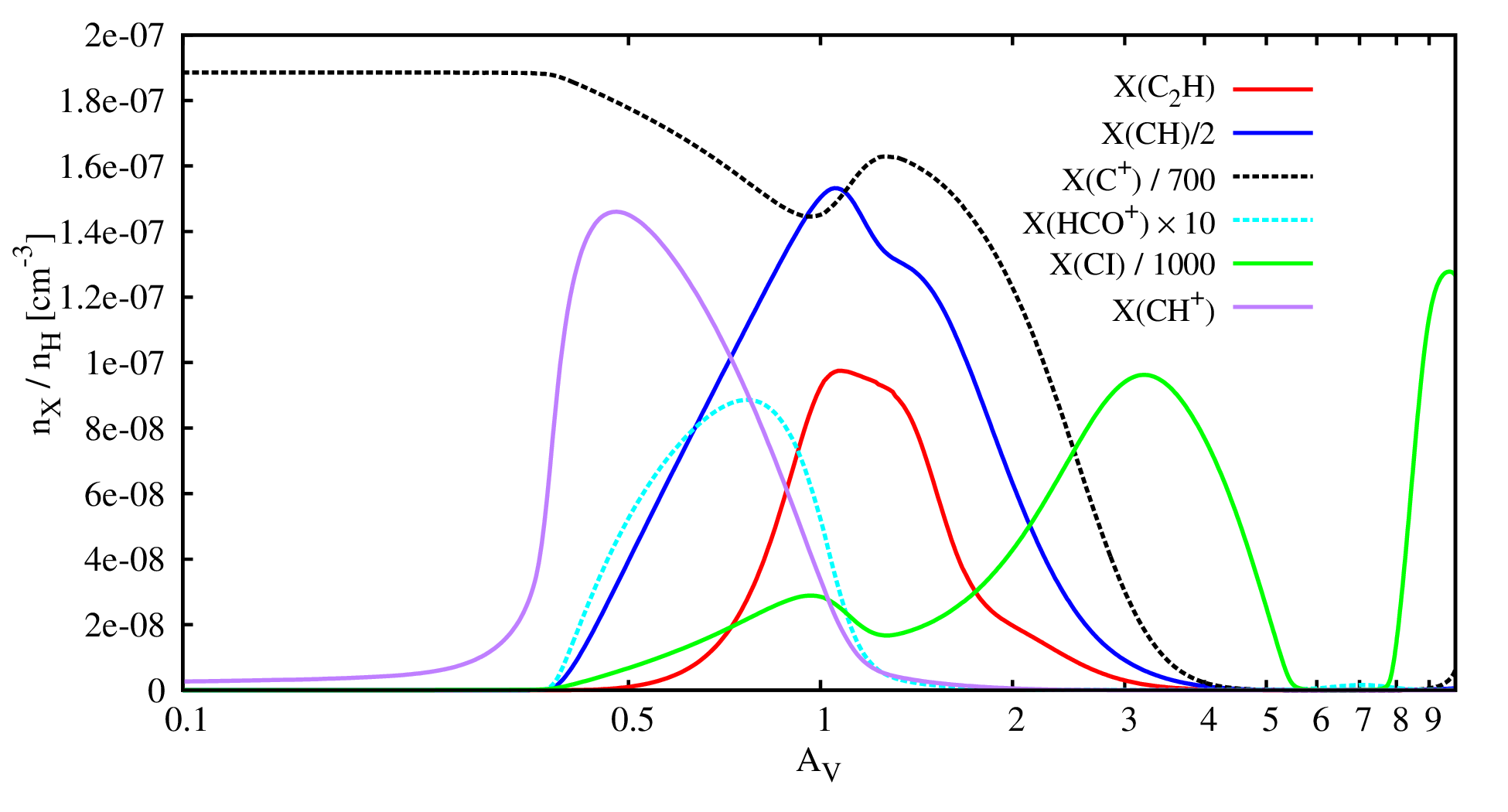}
\caption{The abundances of C$_2$H, CH, C$^+$, HCO$^+$, and [C{\sc{i}}] corresponding to a PDR model with a pressure of $P=10^8$ cm$^{-3}$ K, illuminated by a radiation field of $\chi=10^4$ on the front, and a radiation field of $\chi=10$ on the back side of the cloud.}
\label{abundances_c2hpaper}

\includegraphics[width=4.4cm, angle=-90, trim=0cm 0cm 0cm 0cm,clip=true]{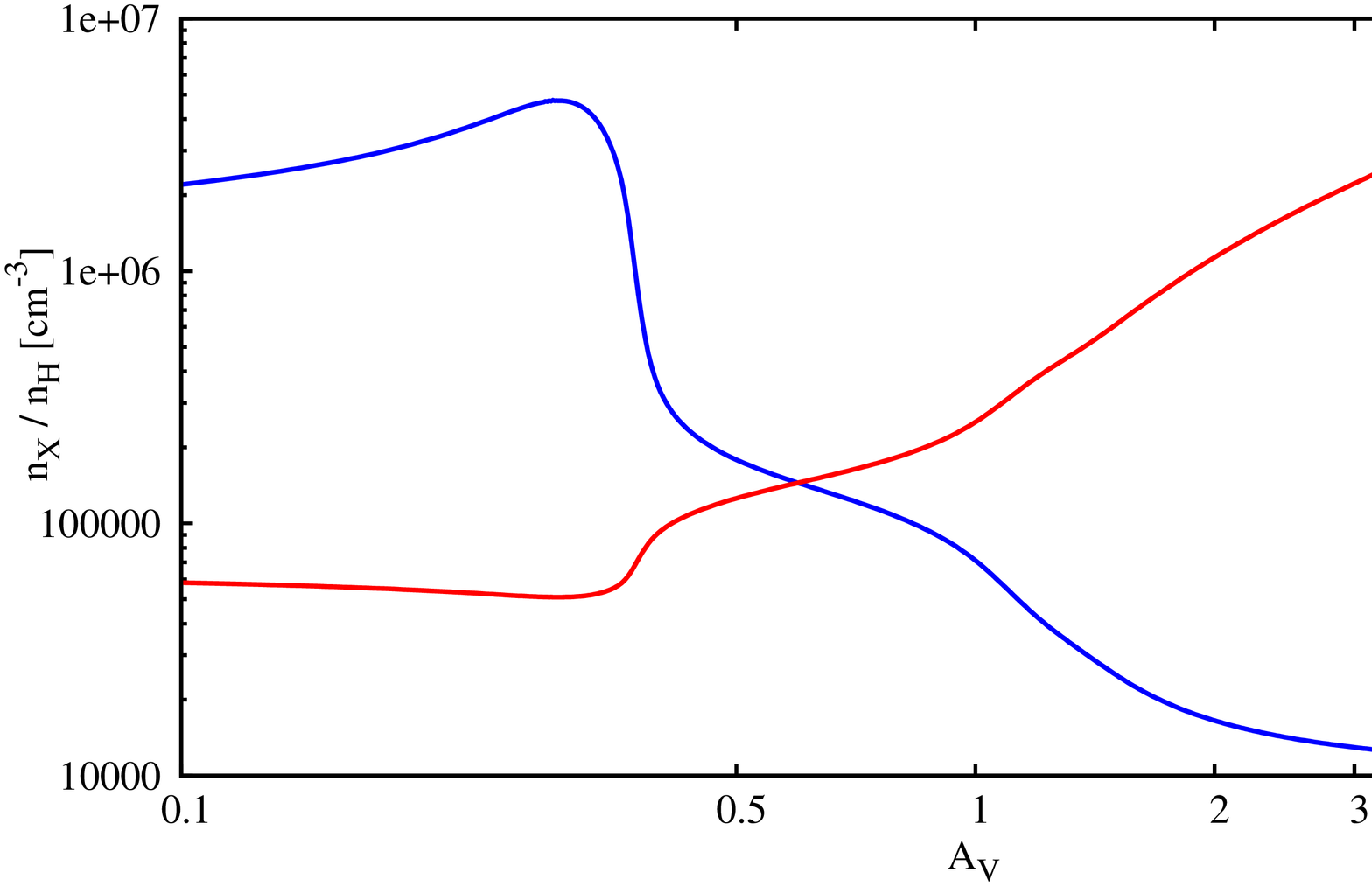}
\caption{The temperature (blue) and density (red) structure corresponding to the isobaric PDR model with a pressure of $P=10^8$ cm$^{-3}$ K, illuminated by a radiation field of $\chi=10^4$ on the front, and a radiation field of $\chi=10$ on the back side of the cloud.}
\label{structure_c2hpaper}
\end{figure}

Figure \ref{abundances_c2hpaper} shows the abundances of C$_2$H, CH, C$^+$, HCO$^+$, [C{\sc{i}}], and CH$^+$ corresponding to the model, as a function of visual extinction up to a visual extinction of 10. Most C$_2$H is located near the surface of the PDR, in a relatively narrow region, at visual extinctions of $A_{\rm{V}}\sim1-2$. The gas temperature in this region predicted by the PDR model is in the range between $\sim$570 K and 140 K, while the H$_2$ volume densities are in the range of 10$^5$ cm$^{-3}$ and $6\times10^5$ cm$^{-3}$, as shown in Fig. \ref{structure_c2hpaper}. 
In the range of $A_V=0.8\dots 1.5$ responsible for the main C$_2$H emission, the PDR model predicts temperatures that cover both components suggested by the RADEX models, 100-150~K and 400~K. 
The H$_2$ volume densities suggested by the RADEX model for the lower-$N$ transitions (10$^5$-10$^6$ cm$^{-3}$) is close to those corresponding to the PDR model.
There are, however, two discrepancies: the RADEX H$_2$ volume density of 5$\times$10$^6$ cm$^{-3}$ for the hot component is above the density range predicted by the model for the C$_2$H emitting gas. Moreover, the temperature-density structure calculated in the isobaric model would predict the highest temperature gas to trace the lowest H$_2$ densities, and the lowest temperature gas to trace the highest H$_2$ densities. This behaviour results in the over-prediction of the intensities of lowest-$N$ transitions and the under-prediction of the intensities of the highest-$N$ transitions. This suggests that a single isobaric model to describe the C$_2$H emitting gas is an over-simplification. In contrast, the PDR is known to be out of equilibrium showing signs of photo-evaporation \citep{stoerzerhollenbach1999}. 
Fig. \ref{abundances_c2hpaper} also shows that the CH$^+$ abundance peaks at lower visual extinctions compared to C$_2$H, and is consistent with gas at higher temperatures and lower H$_2$ densities compared to C$_2$H, which was also confirmed by non-LTE radiative transfer models \citep{nagy2013}. This supports the assumption that most C$_2$H originates in a different gas compared to what CH$^+$ traces.
The dominant C$_2$H formation and destruction processes that the model suggests are shown in Sect. \ref{form_dest}.

The PDR model quoted above is a simplification, but it confirms that C$_2$H toward the Orion Bar originates in more than one pressure component. Based on the physical parameters suggested by the PDR model, the C$_2$H emitting gas is also most likely to trace the edges of dense clumps exposed to UV irradiation close to the ionization front of the PDR.
Moreover, C$_2$H is predicted to be closely confined to the dense surface of the clump near $A_{\rm{V}} \sim 1$. This is consistent with our excitation analysis which argues that this gas is directly heated by the UV photons.

\section{Formation and destruction processes}
\label{form_dest}

Even though we cannot constrain the formation and destruction of C$_2$H with the observations presented in this paper, the PDR models explained above give an indication about the most important processes involved in the chemistry of C$_2$H. For Orion Bar conditions represented by the isobaric model with $P=10^8$ K~cm$^{-3}$ and $\chi=10^4$, the most important processes that contribute in the formation of C$_2$H are
\begin{equation}
\label{rate1}
\mathrm{H_2 + C_2 = C_2H + H}
\end{equation}
\begin{equation}
\label{rate2}
\mathrm{C_2H_2 +} h\nu \mathrm{= C_2H + H}
\end{equation}
\begin{equation}
\label{rate3}
\mathrm{C_2H_2^+ + e^- = C_2H + H}
\end{equation}
\begin{equation}
\label{rate4}
\mathrm{C_3H^+ + e^- = C_2H + C.}
\end{equation}
These processes contribute $\sim$67.1\%, $\sim$27.2\%, $\sim$4.8\%, and $\sim$0.9\% of the total C$_2$H abundance, respectively. These values refer to a depth equivalent to $A_{\rm{V}}\sim1$, which is part of the region, where C$_2$H abundances peak.
An alternative formation route of C$_2$H can be expected in PDRs. Small hydrocarbons, such as C$_2$H can form via the photodissociation of small PAHs ($N_{\rm{C}} \leq 24$) \citep{{uselibacchittajoblin2007}}. This formation route is not included in the Meudon PDR code, however, based on a comparison of the spatial distribution of C$_2$H to that of a Spitzer map at 8 $\mu$m \citet{cuadrado2015} conclude that this formation route cannot dominate the formation of C$_2$H for the Orion Bar.
The most important destruction processes at the same $A_{\rm{V}}\sim1$ are
\begin{equation}
\label{rate5}
\mathrm{C_2H +} h\nu \mathrm{= H + C_2}
\end{equation}
\begin{equation}
\label{rate6}
\mathrm{C_2H + H_2 = H + C_2H_2}
\end{equation}
\begin{equation}
\label{rate7}
\mathrm{C_2H + C^+ = H + C_3^+.}
\end{equation}
\begin{equation}
\label{rate8}
\mathrm{C_2H + O \rightarrow CH + CO}
\end{equation}
These processes are responsible for the destruction of $\sim$68.1\%, $\sim$28.9\%, $\sim$1.5\%, and 0.66\% of C$_2$H, respectively.
C$_2$H abundances therefore peak in a region, where the molecular fraction ($f$=2$N$(H$_2$)/($N$(H)+2$N$(H$_2$))) / H$_2$ density is high enough to produce C$_2$H via reactions between H$_2$ and C$_2$, and the radiation field is low enough, that the C$_2$H production compensates its destruction by FUV photons. With the used model parameters, the molecular fraction is 30\%, the H$_2$ volume density about 10$^5$ cm$^{-3}$, and the gas temperature is around 570 K at $A_{\rm{V}}\sim1$. At a depth equivalent to $A_{\rm{V}}\sim1.5$, where still significant C$_2$H is located, the temperature decreases to about 270 K, the molecular fraction increases to $\sim$50\%, and the H$_2$ volume density is $3\times10^5$ cm$^{-3}$.
The reaction rates are summarized in Table \ref{react_rates}.

\begin{table*}[!ht]
\begin{minipage}[!h]{\linewidth}\centering
\caption{The rate coefficients for the formation and destruction reactions of C$_2$H toward the Orion Bar. For two-body reactions the rates are given by the Arrhenius-type formula of $\alpha \left(\frac{T}{300}\right)^\beta \exp\left(\frac{-\gamma}{T}\right)$ (cm$^3$ s$^{-1}$), and for photo-reactions by the $k = \alpha \exp(-\gamma A_{\rm{V}})$ (s$^{-1}$) formula.
}
\label{react_rates}
\renewcommand{\footnoterule}{}
\renewcommand{\thefootnote}{\alph{footnote}}
\begin{tabular}{llllll}
\hline
Reaction& $\alpha$& $\beta$& $\gamma$& T (K)& Reference\\
\hline
(\ref{rate1})& 1.78$\times$10$^{-10}$& 0& 1469& 295-493& \citet{pitts1982}\\
(\ref{rate2})& 3.30$\times$10$^{-9}$& 0& 2.27& 10-41000& \citet{vandishoeck2006}\\
(\ref{rate3})& 1.35$\times$10$^{-7}$& -0.5& 0& -& \citet{florescumitchellmitchell2006}\\
(\ref{rate4})& 1.50$\times$10$^{-7}$& -0.5& 0& 10-300& \citet{prasadhuntress1980}\\
(\ref{rate5})& 1.60$\times$10$^{-9}$&  0& 2.3& -& \citet{vanhemertvandishoeck2008}, \citet{vandishoeck2006}\\
(\ref{rate6})& 2.90$\times$10$^{-12}$& 1.75& 539&  178-4650& \citet{kruseroth1997}\\
(\ref{rate7})& 1.00$\times$10$^{-9}$& -0.5& 0& 10-41000& \citet{prasadhuntress1980}\\
(\ref{rate8})& 1.00$\times$10$^{-10}$& 0.0& 0& 10-2500& KIDA critical evaluation (based on experimental data)\\
\hline
\end{tabular}
\end{minipage}
\end{table*}

\section{Discussion}

We presented C$_2$H observations toward the Orion Bar corresponding to five rotational transitions, in order to probe the physical parameters of the C$_2$H emitting gas. The observed line intensities are not consistent with a single gas component, but suggest that C$_2$H originates in at least two different components. The total C$_2$H column density of about 10$^{14}$ cm$^{-2}$ is dominated by gas corresponding to a kinetic temperature of 100-150~K, which is consistent with the average value of 85$\pm$30 K derived by \citet{hogerheijde1995} and an H$_2$ volume density of 10$^5$-10$^6$ cm$^{-3}$. The intensities of the higher-$N$ transitions suggest the existence of a higher density (5$\times$10$^6$ cm$^{-3}$) hot ($T\sim400$ K) gas component. This high-pressure gas contributes only a small fraction of the C$_2$H column density, but is required to explain the higher-$N$ C$_2$H transitions. The rotational diagram is also consistent with at least two different gas components.
A high-density gas component toward the Orion Bar, interpreted as photoevaporating dense clumps, was previously found to be traced by OH \citep{goicoechea2011}. The rotational $\Lambda$-doublets detected using \textit{Herschel}/PACS are consistent with warm (160-220~K) gas at densities of $n_{\rm{H}}$=10$^6$-10$^7$ cm$^{-3}$. The high-$N$ rotational transitions of C$_2$H analysed in this paper may trace similar areas of the dense filament surfaces.

Lower-$N$ C$_2$H doublets ($N$=1-0,...,4-3) toward the Orion Bar were analysed by \citet{cuadrado2015}. Based on a rotational diagram analysis and assuming a uniform beam filling factor, they derive C$_2$H column densities of (4.2$\pm$0.2)$\times$10$^{14}$ cm$^{-2}$ and (3.7$\pm$0.6)$\times$10$^{14}$ cm$^{-2}$ toward the dissociation front and the molecular peak of the Orion Bar, about a factor of four above the total C$_2$H column density derived from the high-$N$ transitions presented in this paper. The rotational temperature they find for the $N$=1-0,...,4-3 transitions 26$\pm$1~K, which is significantly below our cold-component temperature of 43$\pm$0.2~K, determined from the $N$=6-5,...,8-7 doublets. 
The combination of the higher column density and lower excitation temperature provides a similar flux around N=5-4 consistent with our measurements, but leading to a continuation of our rotational diagram (Fig. \ref{c2h_rotdiagram}) with a steeper slope towards smaller energies. The low-$N$ observations of \citet{cuadrado2015} therefore also trace somewhat more extended colder gas than our measurements. 
Based on non-LTE radiative transfer models \citet{cuadrado2015} estimate an H$_2$ volume density of $\gtrsim$10$^5$ cm$^{-3}$ and a kinetic temperature of $\gtrsim$150~K. These parameters are close to those that our RADEX models suggest for the bulk of the C$_2$H column density, corresponding mostly to the $N$=6-5,...,8-7 doublets: $n$(H$_2$)=10$^5$-10$^6$ cm$^{-3}$, $T_{\rm{kin}}\sim100-150$~K.

The C$_2$H column density of 10$^{14}$ cm$^{-2}$ can be converted to a C$_2$H abundance using C$^{17}$O transitions observed as part of our HIFI line survey (Nagy et al. 2015, in prep.). The beam-averaged C$^{17}$O column density based on the four observed transitions (from 5-4 to 8-7) based on a rotational diagram is 1.3$\times$10$^{15}$ cm$^{-2}$. This C$^{17}$O column density is equivalent to a CO column density of 2.3$\times$10$^{18}$ cm$^{-2}$, adopting abundance ratios of $^{16}$O/$^{18}$O=560 and $^{18}$O/$^{17}$O=3.2 \citep{wilsonrood1994}. Assuming a CO abundance of 1.1$\times$10$^{-4}$ for the
Orion Bar PDR \citep{johnstone2003} this CO column density is equivalent to an H$_2$ column density of 2.1$\times$10$^{22}$ cm$^{-2}$. 
This value is close to the N(H$_2$)=3$\times$10$^{22}$ cm$^{-2}$ derived by \citet{cuadrado2015} close to the CO$^+$ peak position, based on the C$^{18}$O 1-0, 2-1, and 3-2 transitions. 
This gives a C$_2$H abundance of 5$\times$10$^{-9}$ with respect to H$_2$. \citet{vanderwiel2009} derived a C$_2$H abundance of 2$\times$10$^{-9}$ for a beam size of 15$''$. \citet{fuente1996} derived a C$_2$H abundance of 1.2$\times$10$^{-8}$ with respect to H$_2$ toward the Orion Bar within a 26$''$ beam size.
When comparing to the total hydrogen column density $N_{\rm{H}}=N({\rm{H}})+2N({\rm{H_2}})$ \citet{cuadrado2015} measured C$_2$H abundances of (0.7-2.7)$\times$10$^{-8}$ with respect to H nuclei. Adding the $N({\rm{H}})=3\times10^{21}$ cm$^{-2}$ from \citet{vanderwerf2013} to our H$_2$ column,
gives a C$_2$H abundance of 2.2$\times$10$^{-9}$ with respect to H nuclei.
The higher C$_2$H abundance measured by \citet{cuadrado2015} is due to tracing a cooler component as well using lower-$N$ transitions.

C$_2$H has been detected toward other PDRs with various physical conditions, including Mon R2 (\citealp{rizzo2005}, \citealp{ginard2012}, \citealp{pilleri2013}), the Horsehead PDR (\citealp{teyssier2004}, \citealp{pety2005}), and NGC 7023 \citep{fuente1993}. Similar to what we found for the Orion Bar, multiple C$_2$H transitions measured toward Mon R2 also suggest a two-component fit to describe C$_2$H toward this region \citep{pilleri2013}. However, only one of the two components found toward the Mon R2 region corresponds to a dense highly UV-illuminated PDR, and the other to a low UV-illuminated lower density envelope. The PDR component toward the Mon R2 originates in an H$_2$ volume density of 3$\times$10$^6$ cm$^{-3}$, which is similar to what we found toward the Orion Bar for both components.
Unlike the Orion Bar and Mon R2 PDRs, C$_2$H toward the NGC 7023 region is likely to be related to a low-density envelope of NGC 7023, based on the comparison of the observations and chemical models \citep{fuente1993}.
C$_2$H has also been detected toward the low UV-illumination (60 in \citet{draine1978} units) Horsehead PDR, and was found to be spatially correlated with HCO \citep{gerin2009}.

In addition to PDRs, C$_2$H has also been detected in other types of regions as well. \citet{watt1988} studied the C$_2$H $N$=4$-$3 transition in several molecular clouds. Based on the comparison of the observations and chemical models they conclude that C$_2$H emission arises from dense (10$^4$-10$^5$ cm$^{-3}$) gas, but not from very dense gas with H$_2$ densities of $>$10$^6$ cm$^{-3}$.
\citet{padovani2009} studied the C$_2$H $N$=1-0 and $N$=2-1 transitions toward two prestellar cores, and found evidence of deviations from LTE level populations. 
Deviations from LTE are also important in the case of the transitions studied here for the Orion Bar, as the excitation temperatures derived from the rotational diagrams are well below the kinetic temperatures derived using the non-LTE analysis.
\citet{beuther2008} studied C$_2$H in regions that represent different evolutionary stages of high-mass star formation, including infrared dark clouds, high-mass protostellar objects, and ultracompact H{\sc{ii}} regions. Based on chemical models they suggest that C$_2$H traces the initial conditions of massive star formation. In the outer cloud regions discussed in their models the C$_2$H abundance is high due to the interstellar UV radiation field, which dissociates CO, enlarging the abundance of carbon. These outer cloud regions discussed by \citet{beuther2008} also represent a similar case to PDRs, including the Orion Bar.
Toward the DR21(OH) high-mass star forming region \citet{mookerjea2012} found that the observed C$_2$H and c-C$_3$H$_2$ abundances are consistent with a chemical model with an H$_2$ volume density of 5$\times$10$^6$ cm$^{-3}$, similar to what our RADEX models suggest toward the Orion Bar. Apart from star-forming regions, C$_2$H has also been detected toward several cold dark clouds \citep{wootten1980}. They explain the C$_2$H/HC$_3$N abundance ratio to be consistent with gas-phase chemistry.
C$_2$H has been detected toward planetary nebulae including NGC 7027 \citep{zhang2008}. \citet{zhang2008} suggested C$_2$H to be produced via the photodissociation of C$_2$H$_2$ toward this region.
C$_2$H has also been observed toward diffuse clouds as well (\citealp{{lucasliszt2000}}, \citet{gerin2011}). 
Both \citet{lucasliszt2000} and \citet{gerin2011} suggested a correlation between C$_2$H and c-C$_3$H$_2$ in diffuse clouds. \citet{gerin2011} also suggest C$_2$H to be a tracer of molecular hydrogen in diffuse and translucent gas.

In conclusion, among the other environments mentioned above, C$_2$H toward the Orion Bar is likely to trace a similar gas component to what was found toward Mon R2 and toward DR21(OH).

\section{Summary}

We have presented \textit{Herschel}/HIFI observations of five high-$N$ C$_2$H rotational doublets ($N$=6$-$5,...,10$-$9) toward the CO$^+$ peak of the Orion Bar. 

These observations helped to constrain the excitation of C$_2$H. A single-component rotational diagram of C$_2$H suggests a rotation temperature of 64~K and a C$_2$H column density of $\sim$4$\times$10$^{13}$ cm$^{-2}$. The rotational diagram is also consistent with two different gas components, corresponding to rotation temperatures of 43~K and 123~K, and column densities of $8\times10^{13}$ cm$^{-2}$ and $2\times10^{13}$ cm$^{-2}$ for the three lower-$N$ and for the three higher-$N$ transitions, respectively.

A C$_2$H $N$=6$-$5 HIFI map shows C$_2$H emission along the Orion Bar and perpendicular to it, corresponding to the Orion Ridge. 
Based on the comparison of the C$_2$H integrated intensities to those of CH, HCO$^+$, and [C{\sc{i}}], CH and HCO$^+$ are the best C$_2$H tracers both toward the Orion Ridge and toward the Orion Bar among the molecules studied in this paper. The calculated correlation coefficients toward the Orion Ridge (in the velocity interval of 7-8.5 \kms) are 0.92 and 0.89 for CH and HCO$^+$, respectively. The correlation coefficients toward red-shifted velocities compared to the expected LSR velocity of the Orion Bar (10-11.5 \kms) are 0.88 and 0.89 for CH and HCO$^+$, respectively.
The correlation between C$_2$H and C{\sc{i}} is lower toward both regions, suggesting that C$_2$H and C{\sc{i}} do not entirely trace the same gas.

Based on non-LTE radiative transfer models, the detected C$_2$H line intensities are consistent with C$_2$H to originate at least in two different gas components. One of these gas components dominates the C$_2$H total column density and the intensity of the lower$-N$ transitions, and can be related to warm ($T_{\rm{kin}}\sim100$-150 K) and dense ($n$(H$_2$)$\sim$10$^5$-10$^6$ cm$^{-3}$) gas. The second component adds only a small fraction to the C$_2$H column density, but is required to fit the intensity of the higher-$N$ transitions. This component is related to hot ($T_{\rm{kin}}\sim400$ K) and dense ($n$(H$_2$)$\sim$5$\times$10$^6$ cm$^{-3}$) gas. A simple PDR model representing the Orion Bar with a plane-parallel slab of gas and dust confirms that C$_2$H is likely to be related to more than a single pressure component, unlike CH$^+$ toward the Orion Bar.

Based on the physical parameters derived for the high-$N$ C$_2$H transitions analysed in this paper, the $N$=6-5,...,10-9 C$_2$H doublets may trace the edges of dense clumps exposed to UV radiation near the ionization front of the Orion Bar.

\begin{acknowledgements}
We thank the referee and the editor Malcolm Walmsley for useful comments which helped to improve our manuscript.
We thank Robert Simon for providing us with a C{\sc{i}} 1-0 NANTEN map. 
We also thank Franck Le Petit for discussions related to including the excitation of new species in the Meudon code.
HIFI has been designed and built by a consortium of institutes and university departments from across
Europe, Canada and the US under the leadership of SRON Netherlands Institute for Space Research, Groningen, The Netherlands with major contributions from Germany, France and the US. Consortium members are: Canada: CSA, U.Waterloo; France: CESR, LAB, LERMA, IRAM; Germany: KOSMA,MPIfR,MPS; Ireland, NUIMaynooth; Italy: ASI, IFSI-INAF, Arcetri-INAF; Netherlands: SRON, TUD; Poland: CAMK, CBK; Spain: Observatorio Astronomico Nacional (IGN), Centro de Astrobiolog\'ia (CSIC-INTA); Sweden: Chalmers University of Technology - MC2, RSS \& GARD, Onsala Space Observatory, Swedish National Space Board, Stockholm University - Stockholm Observatory; Switzerland: ETH Z\"urich, FHNW; USA: Caltech, JPL, NHSC.
HIPE is a joint development by the Herschel Science Ground Segment Consortium, consisting of ESA, the NASA Herschel Science Center, and the HIFI, PACS and SPIRE consortia. 
\end{acknowledgements}

\begin{appendix}    

\section{C$_2$H--e$^-$ rate coefficients}\label{elec}

The electronic ground state symmetry of the radical C$_2$H is $^2\Sigma^+$. Each rotational level $N$ is therefore split by the spin-rotation coupling between $N$ and the electronic spin $S=1/2$ so that each rotational level $N$ has two sub-levels given by $j=N\pm1/2$. In addition, owing to the non-zero nuclear spin of the hydrogen atom ($I$=1/2), each fine-structure level is further split into 2 hyperfine levels $F=j\pm 1/2$. The rotational constant of C$_2$H is 1.46~cm$^{-1}$. The fine- and hyperfine-splittings are typically 0.01 and 0.001~cm$^{-1}$, respectively. The dipole moment of C$_2$H is 0.77~D \citep{woon1995}.

There is, to our knowledge, no previous estimate for the electron-impact hyperfine excitation rate coefficients of the C$_2$H radical. Recent R-matrix calculations have considered electron scattering from C$_2$H \citep{harrisontennyson2011} but they concentrated on bound states of the anion and electronic excitation of the neutral. Here, electron-impact hyperfine rate coefficients for C$_2$H were computed using a three step procedure, similar to that employed recently for the radicals OH$^+$ \citep{vandertak2013} and CN \citep{harrison2013}: {\it i)} rotational excitation rate coefficients for the dipolar ($\Delta N=1$) transitions were first computed using the Born approximation; {\it ii)} fine-structure excitation rate coefficients were then obtained, from the Born rotational rates, using the (scaled) Infinite Order Sudden (IOS) approximation and {\it iii)} hyperfine excitation rate coefficients were finally obtained using again the IOS approximation. 
In the case of polar molecules, the long-range electron dipole interaction is well-known to control the rotational excitation, especially at low collision energies \citep{itikawamason2005}. 
Although the C$_2$H dipole is only moderate, the Born approximation is expected to be reasonably accurate for dipolar transitions. In practice Born cross sections were computed for collision energies ranging from 1~meV to 1~eV and rate coefficients were deduced for temperatures from 10 to 1000~K. 
The IOS approximation was employed to derive the fine-structure rate coefficients in terms of the rotational rates for excitation out of the lowest rotational level $N=0$, following the general procedure of \cite{faurelique2012} (see Eqs.~1-3 of \cite{harrison2013}). 
The IOS approximation was also applied to derive the hyperfine rate coefficients (Eqs.~6-8 in \cite{harrison2013}).

The above three step procedure was applied to the first 26 rotational levels of C$_2$H, that is up to the level $(N, j, F)=(25, 24.5, 24)$ which lies 945~cm$^{-1}$ above the ground state (0, 0.5, 0), resulting in 246 collisional transitions. The largest rate coefficients are of the order of $5\times 10^{-7}$cm$^3$s$^{-1}$, i.e. typically four orders magnitude larger than the C$_2$H-He rate coefficients, and they are expected to be accurate to within a factor of 2. The complete set of de-excitation rate coefficients is available on-line from the LAMDA \citep{schoier2005} and BASECOL \citep{dubernet2013} database.

\section{Extrapolation of collision rates}
\label{extrapol}

\citet{spielfiedel2012} only provided collision rates for temperatures up to 100~K and for rotational levels with up to $N=8$ corresponding to an upper level energy of about 150~K, as given in BASECOL. This is clearly insufficient to model our observations as we observed up to $N=10-9$, the gas temperatures in the Orion  Bar are around 150~K and densities of more than $10^6$~cm$^{-3}$ allow for a collisional excitation of much higher levels that will eventually decay radiatively, feeding the rotational ladder from levels above $N=9$. A solution of the rate equations for the Orion Bar conditions thus asks for the inclusion of levels with energies of a few times the kinetic temperature.

\begin{figure} 
\includegraphics[angle=90,width=8.5cm, trim=0cm 0cm 0cm 0cm,clip=true]{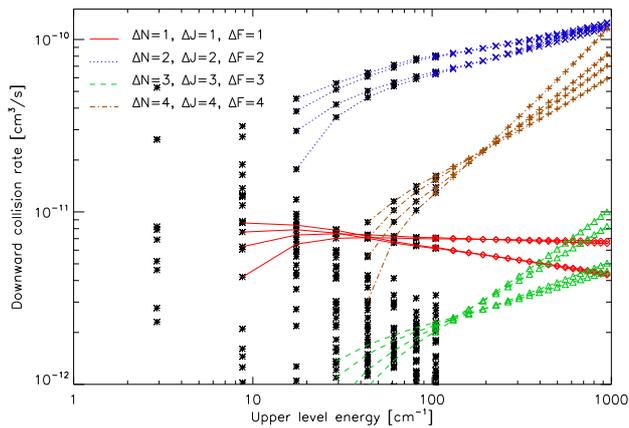} 
\caption{
Rates for the deexcitation of C$_2$H through collision with H$_2$ at 100~K. Black crosses mark all collision rates from \citet{spielfiedel2012} above $10^{-12}$~cm$^3$s$^{-1}$. The curves indicate the different transition series for pure rotational transitions of the molecule, i.e.
$\Delta N =\Delta J = \Delta F$. 
Each series consists of four curves according to the fine and hyperfine structure split of the four $N$ states with approximately the same energy: $J=N\pm/1/2$ and $F=J\pm1/2$.
The continuations of those curves into the coloured symbols show the extrapolation of the rates to the higher level transitions using the formalism from \citet{schoier2005} based on the IOS approximation.}
\label{fig_c2h_extrapol_n}
\end{figure}

Thus we have to extrapolate the rate coefficients from \citet{spielfiedel2012} both in terms of additional levels and higher temperatures. A very general, but numerically demanding approach to extrapolate collision rates was proposed by \citet{neufeld2010}. 
It is applicable to any type of molecules but needs some ''training'' through collision rates from similar species. For C$_2$H we have, however, the advantage that it is basically a linear rotor, only superimposed by the hyperfine interaction. As our measurements show no variation in the hyperfine ratio and the collision rates from \citet{spielfiedel2012} at temperatures above 20~K are dominated by the pure rotational transitions of the linear configuration with $\Delta N =\Delta J = \Delta F$ (see Fig. \ref{fig_c2h_extrapol_n}), it seems justified to concentrate on the rates for these transitions, 
treating C$_2$H as a simple linear rotor.

For such molecules, the IOS approximation is a handy approach to compute the collision rates for any series of $\Delta J$ (see Appendix \ref{elec}) as long as the energy difference of the states is small compared to the kinetic energy of the collision partners. For higher levels, a correction is possible as summarized by \citet{schoier2005}. We use here their Eq.~(13) derived originally by \citet{dejong1975, bieging1998} to extrapolate the collision coefficients to higher levels based on the energy level difference and the behaviour of the collision rates for lower $J$ levels in the series. Due to the additional hyperfine split in C$_2$H we have to perform the fit individually for the different $F-J$ combinations in a pure rotational series. This  assumes that we have no further interaction of the states, as justified by the behaviour of the collision rates ate temperatures between 20 and 100~K. The result is shown in Fig. \ref{fig_c2h_extrapol_n}. The figure shows the collision rate behaviour within the dominant series for the levels computed by \citet{spielfiedel2012} and the extrapolation of the rates up to $N=18$. All other collision rates are much lower so that they are not important for the population of the upper levels.

\begin{figure} 
\includegraphics[angle=90,width=9cm, trim=0.3cm 0cm 0.7cm 0cm,clip=true]{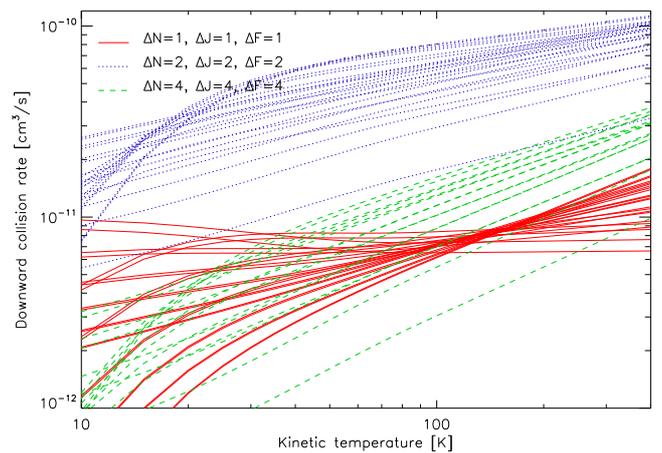} 
\caption{
Temperature dependence for the main collision rates of C$_2$H with H$_2$. The curves are sorted by $\Delta J$ but cover all transitions of the corresponding series. 
Every connected dot in Fig. \ref{fig_c2h_extrapol_n} produces one temperature-dependence graph in this plot. To improve the readability, we have omitted the $\Delta N = \Delta J = \Delta F = 3 $ series here.
Up to 100~K, the lines show the rates computed by \citet{spielfiedel2012}, from 100 to 400~K we extrapolated as a power law with the same slope as measured between 90 and 100~K.}
\label{fig_c2h_extrapol_T}
\end{figure}

Besides the extrapolation to higher level energies, we also need to extrapolate the rates from  \citet{spielfiedel2012} to higher temperatures. To obtain the rate at higher temperatures, we have to compute the integral over collision cross sections for energies of about ten times the kinetic temperature of the gas \citep{flower2001}. This makes the high-temperature computations in principle demanding. However, as we only want to expand the temperature range by a small extend, from 100~K to 400~K, we can use the temperature behaviour measured up to 100~K to estimate the rates above this temperature. The simplest approach would be an extrapolation with the square root of temperature, reflecting the velocity of the collision partners. When inspecting the rate coefficients at temperatures up to 100~K we see, however, that the majority of the rates shows a temperature dependence very different from an exponent of 0.5 (see Fig. \ref{fig_c2h_extrapol_T}). But the double-logarithmic plot shows straight lines for basically all rate coefficients at high temperatures indicating that a power-law extrapolation with a free exponent is a good description of the temperature dependence. The curvature in the curves is very small.

We extrapolated from 100~K to 400~K using the power law determined either between 80 and 100~K or between 90 and 100~K. When comparing the two extrapolations, we found only 11 transitions with a deviation of about 10\,\%. All other transitions showed a better agreement of the rates proving that the power-law description is quite accurate. One should take into account, however, that this approach should not be continued over large dynamic ranges as we clearly expect changing coupling when dealing
with factors of ten or more.

To check the sensitivity of our overall results to the extrapolations, we also ran the RADEX fits performed in Sect. \ref{nonlte} with an alternative collision rate file created just by a constant continuation of the rates from 100~K towards higher temperatures and higher levels. 
For the same input parameters the average excitation temperature then drops from 26.9 K to 24.6 K eventually indicating an overall relatively minor sensitivity of our results to the details of the extrapolations.

We finally note that rate coefficients for the fine-structure excitation of C$_2$H by para-H$_2$($j=0$) have been published recently by \citet{najar2014}. The largest rate coefficients, which correspond to transitions with $\Delta j=\Delta N=2$, were found to differ from those of C$_2$H-He by a factor of $\sim 1.4$, as assumed here. Larger differences were however observed for other transitions, especially those with $\Delta j=\Delta N=1$ (see Fig. 6 in \citealp{najar2014}). Hyperfine-resolved rate coefficients for C$_2$H-H$_2$, for high levels and high temperatures, are therefore urgently required in order to avoid both collider-mass scaling and extrapolations.

\end{appendix}

\end{document}